\documentclass[a4paper,final]{rnti}

\usepackage{a4wide}
\usepackage[frenchb]{babel} 

\usepackage[utf8]{inputenc}

\usepackage{ifthen}
\usepackage{changepage}

\usepackage[x11names,table]{xcolor}
\usepackage[pageanchor=false,colorlinks,plainpages=false]{hyperref}
\hypersetup{%
    pdfauthor={S\'ebastien Nedjar, Fabien Pesci, Rosine Cicchetti, Lotfi Lakhal},
    pdfsubject={Recherche},
    bookmarksnumbered,
    breaklinks=true,
    linktocpage=true,
    pdfstartview={FitH},
    linkcolor=darkgray,
    citecolor=gray,
    urlcolor=blue,
}%

\usepackage{todonotes}

\usepackage{xspace} 
\usepackage{textcomp} 
\usepackage{icomma}

\usepackage{graphicx}
\graphicspath{{Images/}{Figures/}{Experimentations/}}

\usepackage{xcolor}
\usepackage{amsfonts,amsmath,bm}

\usepackage{url}
\usepackage{amssymb}
\usepackage{amsthm}
\usepackage{latexsym}

\usepackage{array}
\usepackage{tabularx}
\usepackage{booktabs}
\usepackage{float}
\usepackage[ruled]{algorithm}
\usepackage{algorithmic}

\usepackage[mathscr]{eucal}

\usepackage{placeins}

\usepackage{footnote}
\makesavenoteenv{table}
\makesavenoteenv{align*}

\usepackage{tikz}
\usepackage{amsmath}
\usetikzlibrary{decorations,arrows,shapes}
\usetikzlibrary{automata}
\usetikzlibrary{positioning}
\usetikzlibrary{shadows}

\DeclareMathOperator*{\Fois}{\bullet}
\DeclareMathOperator*{\Plus}{+}

\newcommand{\accord}{\textsc{Accord}\xspace}
\newcommand{\accords}{\textsc{Accords}\xspace}
\newcommand{\skyline}{\textsc{Skyline}\xspace}
\newcommand{\skylines}{\textsc{Skylines}\xspace}
\newcommand{\skycube}{\textsc{Skycube}\xspace}
\newcommand{\skycubes}{\textsc{Skycubes}\xspace}
\newcommand{\SKY}{\textsc{Sky}\xspace}
\newcommand{\CNA}{\textsc{Cna}\xspace}
\newcommand{\ConceptsAccords}{\textsc{Con\-cepts\-Ac\-cords}\xspace}
\newcommand{\ConceptsSkylines}{\textsc{Con\-cepts\-Sky\-lines}\xspace}

\newcommand{\piSKY}{\ensuremath{\Pi\text{-}\SKY{}}\xspace}

{\end{list}}

\newtheoremstyle{perso}%
  {5pt}{5pt}
  {\normalfont}
  {}
  {\bfseries}{~-}
  { }
  {\thmname{#1}\thmnumber{~#2} \thmnote{\textbf{(#3)}}}

\theoremstyle{perso} 

\newtheorem{definition}{Définition}[section]
\newtheorem{example}{Exemple}[section]
\newtheorem{proposition}{Proposition}[section]
\newtheorem{cor}{Corollaire}[section]

\newtheorem{lemma}{Lemme}[section]
\newtheorem{theorem}{Théorème}[section]

\makeatletter

\newcommand{\algorithmicexit}{\textbf{exit}}
\newcommand{\FORALE}[2][default]{\ALC@it\algorithmicforall\textbf{e}\ #2\ \algorithmicdo\ALC@com{#1}\begin{ALC@for}}
\newcommand{\EXITFOR}{\ALC@it\algorithmicexit\ \algorithmicfor}
\newcommand{\EXITLOOP}{\ALC@it\algorithmicexit\ \algorithmicloop}
\newcommand{\FORALLOL}[2]{\ALC@it\algorithmicforall\ #1\ \algorithmicdo\ #2}
\newcommand{\IFTHENOL}[2]{\ALC@it\algorithmicif\ #1\ \algorithmicthen\ #2 \\}
\newcommand{\IFTHENELSEOL}[3]{\ALC@it\algorithmicif\ #1\ \algorithmicthen\ #2\ \algorithmicelse\ #3}
\newcommand{\THENOLO}[1]{\ALC@it\hspace*{2,15cm}\algorithmicelse\ #1\\}
\newcommand{\THENOLA}[1]{\ALC@it\hspace*{2,65cm}\algorithmicelse\ #1\\}
\newcommand{\THENOLB}[1]{\ALC@it\hspace*{1,85cm}\algorithmicelse\ #1\\}
\newcommand{\THENOLC}[1]{\ALC@it\hspace*{1,95cm}\algorithmicelse\ #1\\}
\makeatother

\floatname{algorithm}{Alg.}

\renewcommand{\algorithmicif}{\textbf{si}}
\renewcommand{\algorithmicthen}{\textbf{alors}}
\renewcommand{\algorithmicelse}{\textbf{sinon}}
\renewcommand{\algorithmicfor}{\textbf{pour}}
\renewcommand{\algorithmicforall}{\textbf{pour tout}}
\renewcommand{\algorithmicdo}{\textbf{faire}}

\renewcommand{\algorithmicloop}{\textbf{boucle}}

\renewcommand{\algorithmicexit}{\textbf{quitter}}

\domaine{E}

\titrecourt{Treillis des concepts \skylines}

\nomcourt{S. Nedjar, %
F. Pesci, %
L. Lakhal \& %
R. Cicchetti}

\titre{%
Treillis des concepts \skylines : %
Analyse multidimensionnelle des \skylines %
fondée sur les ensembles en accord%
}
 
\auteur{S\'ebastien Nedjar, %
Fabien Pesci\footnotemark[1], %
Lotfi Lakhal \& %
Rosine Cicchetti} 
 
\affiliation{%
Laboratoire d'Informatique Fondamentale de Marseille (LIF), CNRS UMR 6166\\
Aix-Marseille Universit\'e,\\
IUT d'Aix-en-Provence, Avenue Gaston Berger, 13625 Aix-en-Provence Cedex
}

\resume{
Le concept de \skyline a été introduit pour mettre en évidence les objets 
«~les meilleurs~» selon différents critères. Une généralisation multidimensionnelle
du \skyline a été proposée à travers le \skycube qui réunit tous les \skylines
possibles selon toutes les combinaisons de critères et permet d'analyser les liens
entre objets \skylines.

Comme le data cube, le \skycube s'avère extrêmement volumineux si bien que des 
approches de réduction sont incontournables. Dans cet article, nous définissons 
une approche de matérialisation partielle du \skycube. L'idée sous-jacente est
d'éliminer de la représentation les Skycuboïdes facilement re-calculables.
Pour atteindre cet objectif de réduction, nous caractérisons un cadre formel : 
le treillis des concepts \accords. Cette structure combine la notion d'ensemble en
accord et le treillis des concepts. À partir de cette structure, nous dérivons le
treillis des concepts \skylines qui en est une instance contrainte.

Le point fort de notre approche est d'être orientée attribut ce qui permet de
borner le nombre de nœuds du treillis et d'obtenir une navigation efficace à travers les 
Skycuboïdes.
}

\summary{
The \skyline concept has been introduced in order to exhibit the best objects according
to all the criterion combinations and makes it possible to analyse the relationships
between \skyline objets.

Like the data cube, the \skycube is so voluminous that reduction approaches are really necessary.
In this paper, we define an approach which partially materializes the \skycube. The underlying idea 
is to discard from the representation the skycuboids which can be computed again easily. 
To meet this reduction objective, we characterize a formal framework: the \textsc{agree} concept 
lattice. From this structure, we derive the \skyline concept lattice which is one of its constrained 
instances.

The strong points of our approach are: (i) it is attribute oriented; (ii) it provides a boundary for 
the number of lattice nodes; (iii) it facilitates the navigation within the Skycuboids.
}

\begin{document}
\pgfdeclarelayer{background}
\pgfdeclarelayer{foreground}
\pgfsetlayers{background,main,foreground}

\footnotetext[1]{Fabien Pesci bénéficie d'une bourse doctorale co-financée par le conseil régional 
\textsc{Paca} et l'entreprise \textsc{Ca2i}.}

\section{Introduction}
Dans un contexte décisionnel, certaines requêtes ne renvoient aucun résultat. Dans ces requêtes, 
l'utilisateur recherche les tuples pour lesquels les valeurs de certains critères sont optimales. 
C'est le caractère «~multicritère~» de ces interrogations qui les rend généralement infructueuses. 
En effet, tel tuple peut être optimal pour un critère mais pas pour un autre, il est alors éliminé 
du résultat alors qu'il aurait pu être pertinent pour l'utilisateur. Par exemple, si l'on considère 
une base de données immobilières, la recherche du logement «~idéal~» peut combiner des conditions 
sur le prix, le plus bas possible, la surface, la plus grande possible, et l'éloignement du lieu de 
travail, le plus réduit possible. Évidemment il est vraisemblable que ce logement idéal n'existe pas,
d'où l'absence de réponse à ce type de requête. Pourtant certains logements pourraient s'avérer 
pertinents pour l'utilisateur parce que, situés dans une zone proche, mais non voisine, ils réunissent 
les critères de surface maximale et de prix minimal.

Afin d'apporter une réponse adéquate au type de requêtes décrites, l'opérateur \skyline~\citep{icde/BorzsonyiKS01}
a été introduit. Il considère l'ensemble des critères de choix d'une recherche comme autant de préférences
et extrait les tuples  globalement optimaux pour cet ensemble de préférences. Ainsi plutôt que de 
rechercher une hypothétique solution idéale, il extrait les candidats les plus proches possibles des
souhaits de l'utilisateur. Son principe général s'appuie sur la notion de dominance. Un objet ou un 
tuple est dit dominé par un autre si, pour tous les critères intéressant le décideur, il est moins 
optimal que cet autre. Un tel tuple est éliminé du résultat, non pas parce qu'il est non pertinent 
pour un des critères mais parce qu'il est non optimal selon la combinaison de tous les critères. En 
d'autres termes, il existe au moins une meilleure solution pour l'utilisateur qui, elle, sera retenue.

De la même manière que le cube de données permet de comprendre les liens existant entre plusieurs 
niveaux d'agrégation, une généralisation multidimensionnelle du \skyline a été proposée à travers 
le \skycube{}~\citep{vldb/YuanLLWYZ05,vldb/PeiJET05}. Cette structure réunit tous les \skylines possibles 
suivant les différentes combinaisons de critères. Il est alors possible de rechercher efficacement des objets 
dominants selon différentes combinaisons de critères. De plus, grâce à cette structure, il devient 
possible d'observer le comportement des objets dominants à travers l'espace multidimensionnel et 
ainsi d'analyser et de comprendre les différents facteurs de dominance. Ce concept étant inspiré du 
cube de données, il souffre des mêmes inconvénients de coût de calcul et d'explosion de l'espace de 
stockage. Ainsi il est naturel, comme pour le data cube, d'essayer d'en proposer des représentations 
réduites et les algorithmes associés.

Dans cet article, nous proposons une approche basée sur le treillis des concepts qui permet de 
matérialiser partiellement les \skycubes{} et donc de réduire la taille de leur représentation tout 
en garantissant la reconstruction complète des résultats. Cette structure combine le concept 
d'ensemble en accord \citep{jacm/BeeriDFS84,jetai/LopesPL02}, issu de la théorie des bases de données, 
et celui du treillis des concepts, fondement de l'analyse de concepts formels. Contrairement à 
\cite{tods/PeiYLJELWTYZ06}, notre approche de réduction est orientée attribut ce qui lui confère 
les mêmes qualités pour la navigation que le \skycube complet.  

Le plan de l'article est le suivant. Au paragraphe 2, nous rappelons le contexte de notre travail : 
le \skycube. Nous formalisons un cadre formel dans lequel s'inscrira notre travail au paragraphe 3. 
Puis nous présentons notre approche de matérialisation partielle des \skycubes.

\section{\skycube pour l'Analyse multidimensionnelle des \skylines}
Dans cette section, nous présentons d'abord l'opérateur \skyline ainsi que la problématique à 
laquelle il répond. Dans un deuxième temps, nous présentons l'analyse multidimensionnelle des 
\skylines à travers le concept de \skycube{}.
\subsection{L'opérateur \skyline}\label{sec:operateur_skyline}
Avant de définir de manière formelle l'opérateur \skyline, il est important de bien situer le contexte 
dans lequel la problématique se pose. En effet, il ne s'applique pas à n'importe quelle 
relation. Pour qu'il puisse effectuer les comparaisons nécessaires, il faut que les différents 
domaines sur lesquels sont définis les attributs, critères de choix de l'utilisateur, soient 
totalement ordonnés~\citep{icde/BorzsonyiKS01,tods/PeiYLJELWTYZ06}.
Par simplicité, les attributs ont systématiquement des valeurs numériques dans nos exemples.

Les tuples de nos relations pouvant être utilisées par l'opérateur \skyline sont de la forme 
$t = (d_1,d_2,\ldots,d_k,c_1,c_2,\ldots,c_l)$ où les $d_i$ sont les dimensions non utilisées par l'opérateur 
\skyline alors que les $c_i$ sont les critères sur lesquels l'utilisateur se fonde pour porter 
son choix.

\begin{table}
\centering
\begin{tabular}{c|ccc|cccc} \toprule
RowId\footnote{Le \textit{RowId} est un attribut implicite dont la valeur sert d'identifiant unique 
à chaque tuple. $t_i$ est le tuple ayant $i$ pour \textit{RowId}.}
      & Propriétaire & ... & Ville      & Prix   & Éloignement & Consommation & Voisins \\ \midrule
 1    & Dupont       &     & Marseille  & 220 &  15         & 275          & 5       \\
 2    & Dupond       &     & Paris      & 100 &  15         & 85           & 1       \\
 3    & Martin       &     & Marseille  & 220 &  7          & 180          & 1       \\
 4    & Sanchez      &     & Aubagne    & 340 &  7          & 85           & 3       \\
 5    & Durand       &     & Paris      & 100 &  7          & 180          & 1       \\
 \bottomrule
 \end{tabular}
\caption{La relation \textsc{Logements}\label{tab:typique}}
\end{table}

\begin{example}
La relation illustrée par la table~\ref{tab:typique} est typique pour l'utilisation du \skyline.
Elle répertorie différents logements. Les attributs dimensions classiques sont ici \emph{Propriétaire} et 
\emph{Ville} et les critères de choix pour trouver le «~meilleur logement~» sont :
le \emph{Prix} de vente en milliers d'euros,
l'\emph{Éloignement} par rapport au lieu de travail en kilomètres,
la \emph{Consommation} Énergétique en kilowattheures par an et par mètre carré,
le nombre de \emph{Voisins}.
\end{example}

\begin{definition}[Relation de dominance]\index{Relation de dominance}
Soit $\mathcal{C} = \{ c_1, c_2, \ldots, c_d \}$ l'ensemble des critères sur lesquels porte l'opérateur 
\skyline\footnote{Sans perte de généralité, nous considérons uniquement le cas où tous les critères %
doivent être minimisés.}. Soit deux tuples $t$ et $t'$, la relation de dominance suivant l'ensemble de 
critères $\mathcal{C}$ est définie comme suit :
\[
  t \succeq_\mathcal{C} t'\; 
  \Leftrightarrow\; 
  t[c_1] \leq t'[c_1] \; et \; 
  t[c_2] \leq t'[c_2] \; et \; 
  \ldots \; et \; 
  t[c_d] \leq t'[c_d]
\]
Lorsque $t \succeq_\mathcal{C} t'$ et $\exists\ c_i \in \mathcal{C}$ tel que $t[c_i] < t'[c_i]$, 
la dominance est stricte, elle est notée $t \succ_\mathcal{C} t'$.
\end{definition}

Lorsqu'un tuple $t$ domine un tuple $t'$ (\emph{i.e.} $t \succeq_\mathcal{C} t'$), cela signifie que 
$t$ est équivalent ou «~meilleur~» que le tuple $t'$ pour tous les critères choisis. Comme nous 
considérons que les critères sont minimisés, les valeurs de $t$ pour tous les critères sont 
inférieures ou égales à celles de $t'$. Ainsi dans le cadre d'une recherche multicritère les tuples 
dominés par d'autres (au moins un) ne sont pas pertinents et sont éliminés du résultat par l'opérateur 
\skyline.

\begin{definition}[L'opérateur \skyline]
Soit $r$ une relation, le \skyline de $r$ suivant $\mathcal{C}$ est l'ensemble des tuples qui ne 
sont dominés par aucun autre, suivant l'ensemble de critères $\mathcal{C}$ :
\[SKY_\mathcal{C}(r) = \{ t \in r \; | \; \nexists \; t' \in r \smallsetminus t, \; t' \succ_\mathcal{C} t \}\]
\end{definition}

\begin{example}
Avec notre relation exemple (\emph{cf.} Table~\ref{tab:typique}) et les critères suivants 
$\mathcal{C} = \{$\emph{Éloignement, Prix}$\}$\footnote{À partir de maintenant, pour simplifier les
notations et lorsque qu'il n'y a aucune ambigüité, nous écrivons les ensembles sans accolades ni 
virgule. Par exemple $\{A,B\}$ est noté $AB$.}, $SKY_\mathcal{C}($\textsc{Logement}$) = \{ t_5 \}$ 
car le tuple $t_5$ domine tous les autres. Il est donc le meilleur logement possible pour l'utilisateur.
\end{example}

\subsection{\skycube{}}
L'opérateur \skyline est un outil fondamental pour l'analyse multicritère des bases de données.
Nous pouvons calculer un \skyline suivant un ensemble de critères définis par l'utilisateur, 
mais lorsqu'il s'agit d'en calculer plusieurs sur les mêmes données, aucun d'eux ne sait exploiter 
à son avantage les liens qui peuvent exister entre les différents \skylines. C'est pour cela 
qu'une structure nommée \skycube{} a été introduite~\citep{vldb/YuanLLWYZ05,vldb/PeiJET05}. On peut dire 
que le \skycube{} est au \skyline ce que le cube de données est au \textsc{Group-By} : 
une généralisation multidimensionnelle. 
Ainsi, dès lors que l'on souhaite répondre rapidement à toutes les requêtes posées sur un \skycube{}, 
il vaut mieux opter pour une pré-matérialisation de ce cube.


\begin{definition}[\skycube{}]
Un \skycube{} est l'ensemble de tous les \skylines dans tous les sous-espaces non vides possibles 
de $\mathcal{C}$ : 
\[
SKYCUBE(r, \mathcal{C}) = \{ (C, SKY_{C}(r)) \mid C \subseteq \mathcal{C}\}
\]
$SKY_{C}(r)$ est appelé le cuboïde \skyline (ou Skycuboïde) du sous-espace $C$. 
Par convention le cuboïde selon l'ensemble de critère vide est vide (\emph{i.e.} $SKY_{\emptyset}(r) = \emptyset$).
\end{definition}

La structure du \skycube{} peut être représentée par un treillis semblable à celui utilisé pour le 
cube de données (\emph{cf.} figure~\ref{skycube_lattice_complet}). Les cuboïdes du \skycube{} 
sont regroupés par niveau en fonction de leur nombre de critères. Ces niveaux sont numérotés en 
partant du bas du treillis (cuboïdes portant sur un seul critère) et en remontant vers le sommet 
(cuboïde suivant tous les critères possibles). 

\begin{example}[\skycube{}]
La figure~\ref{skycube_lattice_complet} donne le \skycube{} associé à la relation \textsc{Logement} 
(\emph{cf.} Table~\ref{tab:typique}). Les critères seront symbolisés par leur initiale.
\end{example}

\newlength{\niveauZero}
\newlength{\niveauUn}
\newlength{\niveauDeux}
\newlength{\niveauTrois}
\newlength{\niveauQuatre}
\newlength{\niveauCinq}

\setlength{\niveauZero}{0bp}
\setlength{\niveauUn}{40bp}
\setlength{\niveauDeux}{90bp}
\setlength{\niveauTrois}{150bp}
\setlength{\niveauQuatre}{190bp}
\setlength{\niveauCinq}{1.5em}

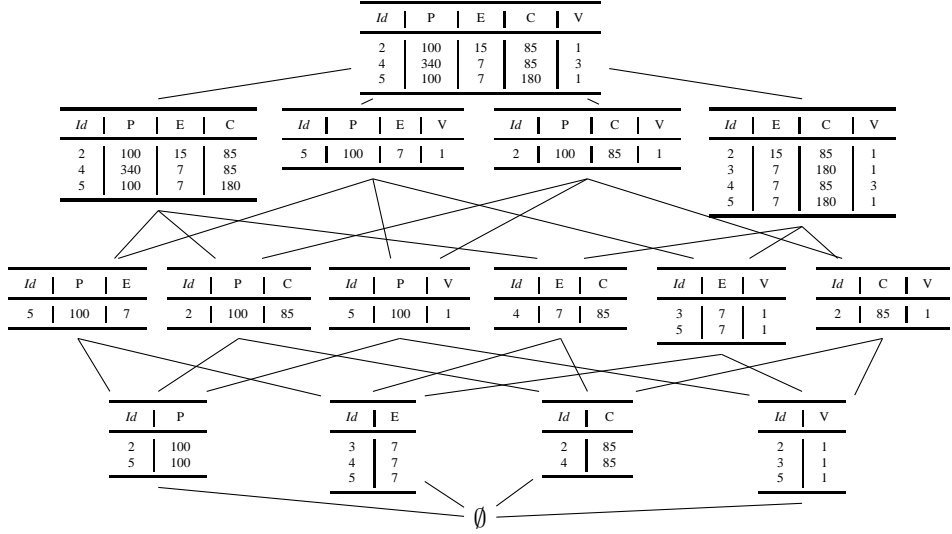
\begin{figure}
\centering
  \begin{tikzpicture}[>=latex,line join=bevel,node distance=0.1cm]

    \node (bottom) at (150bp,\niveauZero) [anchor=north,draw,draw=none] 
          {$\emptyset$};
    
    \node (p1)     at (30bp ,\niveauUn) [anchor=north,draw,draw=none] 
          {\tiny
            \begin{tabular}{c|c} \toprule
              \textit{Id}&P      \\ \midrule 
              2          &100 \\
              5          &100 \\ \bottomrule
            \end{tabular}
          };
   \node (p2) at (110bp,\niveauUn) [anchor=north,draw,draw=none] 
          {\tiny
            \begin{tabular}{c|c} \toprule
              \textit{Id}&E   \\ \midrule 
              3          & 7  \\
              4          & 7  \\
              5          & 7  \\ \bottomrule
            \end{tabular}
          };
    \node (p3) at (190bp,\niveauUn) [anchor=north,draw,draw=none] 
          {\tiny
            \begin{tabular}{c|c} \toprule
              \textit{Id}&C   \\ \midrule 
              2          & 85 \\
              4          & 85 \\ \bottomrule

            \end{tabular}
          };
    \node (p4) at (270bp,\niveauUn) [anchor=north,draw,draw=none] 
          {\tiny
            \begin{tabular}{c|c} \toprule
              \textit{Id}& V \\ \midrule 
              2          & 1 \\
              3          & 1 \\
              5          & 1 \\ \bottomrule
            \end{tabular}
          };

    \node (p5)     at (0bp,\niveauDeux) [anchor=north,draw,draw=none] 
          {\tiny
            \begin{tabular}{c|c|c} \toprule
              \textit{Id}&P      &E  \\ \midrule 
              5          &100 & 7 \\ \bottomrule
            \end{tabular}
          };

    \node (p6) at (60bp,\niveauDeux) [anchor=north,draw,draw=none] 
          {\tiny
            \begin{tabular}{c|c|c} \toprule
              \textit{Id}&P     &C   \\ \midrule 
              2          &100& 85 \\ \bottomrule

            \end{tabular}
          };
          
    \node (p7) at (120bp,\niveauDeux) [anchor=north,draw,draw=none] 
          {\tiny
            \begin{tabular}{c|c|c} \toprule
              \textit{Id}&P      & V \\ \midrule 
              5          &100 & 1 \\ \bottomrule
            \end{tabular}
          };
    
    \node (p8) at (180bp,\niveauDeux) [anchor=north,draw,draw=none] 
          {\tiny
            \begin{tabular}{c|c|c} \toprule
              \textit{Id}&E   &C   \\ \midrule 
              4          & 7  & 85 \\ \bottomrule

            \end{tabular}
          };
    \node (p9) at (240bp,\niveauDeux) [anchor=north,draw,draw=none] 
          {\tiny
            \begin{tabular}{c|c|c} \toprule
              \textit{Id}&E   & V \\ \midrule 
              3          & 7  & 1 \\
              5          & 7  & 1 \\ \bottomrule
            \end{tabular}
          };
    
    \node (p10) at (300bp,\niveauDeux) [anchor=north,draw,draw=none] 
          {\tiny
            \begin{tabular}{c|c|c} \toprule
              \textit{Id}&C    & V \\ \midrule 
              2          & 85  & 1 \\ \bottomrule

            \end{tabular}
          };

    \node (p11)     at (30bp,\niveauTrois) [anchor=north,draw,draw=none] 
          {\tiny
            \begin{tabular}{c|c|c|c} \toprule
              \textit{Id}&P      &E   &C   \\ \midrule 
              2          &100 & 15 & 85 \\
              4          &340 & 7  & 85 \\
              5          &100 & 7  & 180\\ \bottomrule
            \end{tabular}
          };

    \node (p12) at (110bp,\niveauTrois) [anchor=north,draw,draw=none] 
          {\tiny
            \begin{tabular}{c|c|c|c} \toprule
              \textit{Id}&P      &E   & V \\ \midrule 
              5          &100 & 7  & 1 \\ \bottomrule
            \end{tabular}
          };
          
    \node (p13) at (190bp,\niveauTrois) [anchor=north,draw,draw=none] 
          {\tiny
            \begin{tabular}{c|c|c|c} \toprule
              \textit{Id}&P      &C    & V \\ \midrule 
              2          &100 & 85  & 1 \\ \bottomrule

            \end{tabular}
          };
    
    \node (p14) at (270bp,\niveauTrois) [anchor=north,draw,draw=none] 
          {\tiny
            \begin{tabular}{c|c|c|c} \toprule
              \textit{Id}&E   &C    & V \\ \midrule 
              2          & 15 & 85  & 1 \\
              3          & 7  & 180 & 1 \\
              4          & 7  & 85  & 3 \\
              5          & 7  & 180 & 1 \\ \bottomrule
            \end{tabular}
          };

    \node (top)    at (150bp,\niveauQuatre) [anchor=north,draw,draw=none] 
          {\tiny
            \begin{tabular}{c|c|c|c|c} \toprule
              \textit{Id}&P      &E   &C    & V \\ \midrule 
              2          &100 & 15 & 85  & 1 \\
              4          &340 & 7  & 85  & 3 \\
              5          &100 & 7  & 180 & 1 \\ \bottomrule
            \end{tabular}
          };
    \draw [] (bottom) -- (p1.south);
    \draw [] (bottom) -- (p3);
    \draw [] (bottom) -- (p2);    
    \draw [] (bottom) -- (p4.south);
    \draw [] (p1.135) -- (p5.south);
    \draw [] (p1.90) -- (p6.south);
    \draw [] (p1.45) -- (p7.south);
    \draw [] (p2.135) -- (p5.south);
    \draw [] (p2.90) -- (p8.south);
    \draw [] (p2.45) -- (p9.south);
    \draw [] (p3.135) -- (p6.south);
    \draw [] (p3.90) -- (p8.south);
    \draw [] (p3.45) -- (p10.south);    
    \draw [] (p4.135) -- (p7.south);
    \draw [] (p4.90) -- (p9.south);
    \draw [] (p4.45) -- (p10.south);
    \draw [] (p5) -- (p11.south);
    \draw [] (p5.45) -- (p12.south);
    \draw [] (p6.120) -- (p11.south);
    \draw [] (p6.60) -- (p13.south);    
    \draw [] (p7) -- (p12.south);
    \draw [] (p7.45) -- (p13.south);
    \draw [] (p8.120) -- (p11.south);
    \draw [] (p8.60) -- (p14.south);
    \draw [] (p9.120) -- (p12.south);
    \draw [] (p9.60) -- (p14.south);
    \draw [] (p10.135) -- (p13.south);
    \draw [] (p10) -- (p14.south);
    \draw [] (p11.north) -- (top);
    \draw [] (p12.north) -- (top);
    \draw [] (p13.north) -- (top);
    \draw [] (p14.north) -- (top);
    
  \end{tikzpicture}
\caption{Représentation sous forme de treillis du \skycube{} de la relation \textsc{Logement}\label{skycube_lattice_complet}}
\end{figure}

Une requête \skyline multidimensionnelle retourne le sous-ensemble de tuples de la relation 
originelle formant le \skyline dans un sous-espace donné. Clairement, une fois le \skycube{} 
calculé, il est possible de répondre à toute requête efficacement.

\subsubsection{Problèmes associés à l'analyse multidimensionnelle des \skylines}

En général, si un tuple $t$ est dans les \skylines des sous-espaces $C_1$ et 
$C_2$ tels que $C_1 \subset C_2$, pouvons-nous affirmer que $t$ 
appartiendra aussi au \skyline de n'importe quel sous-espace $C$ situé entre 
$C_1$ et $C_2$ ($C_1 \subset C \subset C_2$) ? 
Une telle propriété serait fort attrayante puisqu'elle pourrait considérablement simplifier la 
détermination des \skylines multidimensionnels. Malheureusement elle n'est pas vérifiée dans le 
cas général.

\begin{example}
Avec la relation \textsc{Logement}, comme le montre la figure~\ref{skycube_lattice_complet},
en considérant les \skylines selon (\emph{Prix, Éloignement,Voisin}) et (\emph{Éloignement,Voisin}) on a :
(\emph{Prix, Éloignement,Voisin}) $\supseteq $ (\emph{Éloignement,Voisin}) et 
$SKY_{PEV}($\textsc{Logement}$) \subseteq SKY_{EV}($\textsc{Logement}$)$.

En revanche, si l'on regarde les \skylines suivants (\emph{Prix, Éloignement, Consommation, Voisins}) 
et (\emph{Prix, Éloignement, Consommation}) on a 
(\emph{Prix, Éloignement, Consommation, Voisins}) $\supseteq $  (\emph{Prix, Éloignement, Consommation}) et 
$SKY_{PECV}($\textsc{Logement}$) \supseteq SKY_{PEC}($\textsc{Logement}$)$.
\end{example}

L'observation mise en évidence par cet exemple est la suivante : l'appartenance à un \skyline 
n'est pas monotone, c'est-à-dire qu'un tuple $t$ appartenant à un cuboïde $SKY_{\mathcal{U}}(r)$ 
n'est pas automatiquement contenu dans les ancêtres de ce cuboïde. 

Comme dans le cas du cube de données, le \skycube peut contenir des informations superflues.  
C'est cette problématique qui a motivé la proposition de représentations réduites du \skycube 
faite par~\cite{tods/PeiYLJELWTYZ06}. Notre contribution s'intéresse à la même problématique de 
réduction en combinant nous aussi l'analyse de concepts formels~\citep{book/GanterW99} et le 
\skyline. Pour éviter le coût important de reconstruction des cuboïdes \skylines induit par le 
regroupement orienté valeur de~\cite{tods/PeiYLJELWTYZ06}, notre méthode de réduction choisit une 
approche de regroupement orientée critère en se basant sur les ensembles en accord.

\section{Treillis des Concepts \accords d'une relation}
Dans ce paragraphe, notre objectif est de définir un cadre de travail formel combinant le concept 
d'ensemble en accord et celui de treillis des concepts. Nous proposons une nouvelle structure, le 
treillis des concepts \accords d'une relation, sur laquelle s'appuie notre matérialisation partielle 
du \skycube. Après un rappel des notions d'ensemble en accord et des classes d'équivalence associées, 
nous caractérisons rigoureusement le treillis des concepts \accords.  
\subsection{Ensembles en Accords}
Le concept d'ensemble en accord (ainsi que le système de fermeture associé), introduit 
par~\cite{jacm/BeeriDFS84} pour caractériser la relation d'Armstrong, a été utilisé avec succès pour 
l'extraction de dépendances fonctionnelles exactes et approximatives~\citep{jetai/LopesPL02}.
Deux tuples sont en accord sur un ensemble d'attributs $X$ s'ils partagent la même valeur pour $X$.

\begin{definition}[Ensembles en accord]
Soit $t_i$, $t_j$ deux tuples de $r$ et $X \subseteq \mathcal{C}$ un ensemble d'attributs 
(critères dans notre contexte).
$t_i$, $t_j$ sont en accord sur $X$ si et seulement si $t_i[X] = t_j[X]$. 
L'ensemble des attributs en accord de $t_i$ et $t_j$ est défini comme suit :
\[
\textsc{Acc}(t_i, t_j) = \{C \in \mathcal{C} \mid  t_i[C] = t_j[C] \}
\]
Cette définition peut être généralisée pour un ensemble de tuples $T \subseteq r$ composé d'au moins 
deux éléments :
\[
\textsc{Acc}(T) = \{C \in \mathcal{C} \mid t[C] = t'[C],\ \forall t, t' \in T \}
\]
\end{definition}

\begin{definition}[Ensembles en accord d'une relation]
L'ensemble des attributs en accord d'une relation $r$ est défini comme suit : 
\[
\accords(r) = \{\textsc{Acc}(t_i, t_j) \mid t_i, t_j \in r \text{ et } i \neq j\}
\]
Cet ensemble peut être redéfini de manière équivalente :
\[
\accords(r) = \{\textsc{Acc}(T) \mid \forall T \subseteq r \text{ et } |T| \geq 2\}
\] 
\end{definition}

\begin{example}
Avec la relation \textsc{Logement}, $\textsc{Acc}(t_2, t_5) =PV$
car ces deux tuples partagent la même valeur sur les attributs \emph{Prix}, \emph{Voisins} et ont des 
valeurs différentes pour \emph{Éloignement}, \emph{Consommation}.
De même, $\textsc{Acc}(\{t_3, t_4, t_5\}) =E$ car ces trois tuples ont la même valeur uniquement 
pour le critère \emph{Éloignement}. L'ensemble des ensembles d'attributs en accord de la relation \textsc{Logement} 
est le suivant :\\
$\accords($\textsc{Logement}$)=\{\emptyset, E, P, V, C, PV, ECV \}$.
\end{example}

\begin{definition}[Classe d'équivalence d'un tuple] 
Soit $r$ une relation et $C \subseteq \mathcal{C}$ un ensemble de critères. La classe d'un tuple $t$ 
selon $C$, notée $[t]_C$, est définie comme l'ensemble des identifiants $i$ ($Rowid$) de tous les 
tuples $t_i\in r $ en accord avec $t$ selon $C$ (\emph{i.e.} l'ensemble des identifiants des tuples
$t_i$ partageant avec $t$ les mêmes valeurs pour $C$). Nous avons donc :
\[
[t]_C = \{i \in Tid(r) \mid t_i[C] = t[C]\}
\]
\end{definition}

\begin{example}
Avec la relation \textsc{Logement}, $[t_2]_{P} =\{2, 5\}$ car seuls les tuples $t_2$ et $t_5$ partagent 
la même valeur sur le critère \emph{Prix}.
\end{example}

%
%

\subsection{Concepts \accords d'une relation}
Notre objectif est de définir un treillis des concepts particulier se basant sur les ensembles en 
accord et les partitions~\citep{tods/Spyratos87}. Pour atteindre ce but, nous caractérisons une 
instance de la connexion de Galois entre d'une part le treillis des parties de l'ensemble des 
critères et d'autre part le treillis des partitions de l'ensemble des identifiants de tuples. 
Cette connexion nous permet de dériver des opérateurs de fermeture duaux, de définir un concept 
\accord et de caractériser le treillis des concepts \accords.

\begin{definition}\label{def:f_g}
  Soit $Rowid: r \rightarrow \mathbb{N}$ une application qui associe à chaque tuple un unique entier 
  naturel et $Tid(r) = \{Rowid(t) \mid t \in r\}$. Soit $f$, $g$ deux applications entre 
  les ensembles ordonnés $\langle\Pi(Tid(r)), \sqsubseteq \rangle$\footnote{La définition du %
   treillis des partitions $\Pi(E)$ d'un ensemble $E$ est rappelée en annexe.} et 
  $\langle\mathscr{P}(\mathcal{C}), \subseteq\rangle$ qui sont définies comme suit :
  \[
    \begin{array}{lrcl}
      f:&	\langle\Pi(Tid(r)), \sqsubseteq \rangle &
      \longrightarrow & 
      \langle\mathscr{P}(\mathcal{C}), \subseteq\rangle\\
      
      & \pi & \longmapsto & \displaystyle\bigcap_{[t] \in \pi} \textsc{Acc}(\{t_i \mid i \in [t]\}) \\

      g:&	\langle\mathscr{P}(\mathcal{C}), \subseteq\rangle &
      \longrightarrow & 
      \langle\Pi(Tid(r)), \sqsubseteq \rangle\\
      
      & C & \longmapsto & \displaystyle \{[t]_{C} \mid t \in r \} \\
    \end{array}
    \]
\end{definition}

Pour un ensemble de critères $C$, l'ensemble des classes d'équivalence selon $C$ forme une 
partition de $Tid(r)$. C'est la fonction $g$ qui associe à $C$ cette partition des identifiants. 
Cette dernière est notée $\pi_C$ et définie comme suit : $\pi_C = g(C)$. L'ensemble de toutes les 
partitions $\pi_C$ possibles est noté $\Pi_{\mathscr{P}(\mathcal{C})}$. La fonction $f$ fait 
l'association inverse de $g$.

\begin{example}\label{ex:f_g}
Avec la relation \textsc{Logement}, en prenant les ensembles de critères $E$, $EC$, $ECV$ et $PV$, 
on a $g(E) = \{12,345\}$\footnote{Sont dans la même classe d'équivalence d'une part $t_1$, $t_2$ et %
d'autre part $t_3$, $t_4$, $t_5$.}, $g(EC) = \{1,2,35,4\}$, $g(ECV) = \{1,2,35,4\}$ et 
$g(PV) = \{1,25,3,4\}$. Avec les partitions $\{1,2,35,4\}$ et $\{1,2,345\}$, on a 
$f(\{1,2,35,4\}) = ECV$ et $f(\{1,2,345\}) = E$.
\end{example}

\begin{proposition}
Le couple d'applications $gc = (f,\ g)$ est une connexion de Galois entre le treillis des parties 
des critères $\mathcal{C}$ et le treillis des partitions de $Tid(r)$.
\end{proposition}
   
\begin{definition}[Opérateurs de fermeture]
Le couple $gc = (f,\ g)$ étant un cas particulier de la connexion de Galois, les compositions 
$f \circ g$  et $g \circ f$  des deux applications sont des opérateurs de 
fermeture~\citep{book/GanterW99} définis ci-dessous.
  \[
    \begin{array}{lrcl}
      h:&	\mathscr{P}(\mathcal{C})& \longrightarrow & \mathscr{P}(\mathcal{C})\\
        & C & \longmapsto & \displaystyle f( g (C) ) = \bigcap_{\substack{C'\in \accords(r)\\
                                                                              C \subseteq C'}} C' \\
                                                                              
      h':&	\Pi(Tid(r)) & \longrightarrow & \Pi(Tid(r))\\
      & \pi & \longmapsto & \displaystyle g( f (\pi) ) = 
      \Fois_{\substack{\pi'\in \Pi_{\mathscr{P}(\mathcal{C})}\\
                       \pi \sqsubseteq \pi'}} \pi' \\
    \end{array}
    \]
\end{definition}

\begin{cor}
$h$ et $h'$ satisfont les propriétés suivantes:
 \begin{enumerate}
 \item $C \subseteq C' \Rightarrow h(C) \subseteq h(C')$ et  
       $\pi \sqsubseteq \pi' \Rightarrow h'(\pi) \sqsubseteq h'(\pi')$ (isotonie)
 \item  $C \subseteq h(C)$ et $\pi \sqsubseteq h'(\pi)$ (extensivité)
 \item $h(C) = h(h(C))$ et $h'(\pi) = h'(h'(\pi))$ (idempotence)
\end{enumerate}
\end{cor}

\begin{example}\label{ex:fermeture}
Avec la relation \textsc{Logement}, en prenant les ensembles de critères $EC$ et $ECV$, d'après 
l'exemple précédent on a :
\begin{itemize}
	\item $h(EC) = f(g(EC)) = f(\{1,2,35,4\}) = ECV$  
	\item $h(ECV) = f(g(ECV)) = f(\{1,2,35,4\}) = ECV$
\end{itemize}
Avec les partitions $\{1,2,35,4\}$ et $\{1,2,345\}$, on a :
\begin{itemize}
	\item $h'(\{1,2,35,4\}) = g(f(\{1,2,35,4\})) = g(ECV) = \{1,2,35,4\}$
  \item $h'(\{1,2,345\}) = g(f(\{1,2,345\})) = g(E) = \{12,345\}$
\end{itemize}
\end{example}

\begin{definition}[Concepts \accords]
Un concept \accord d'une relation $r$ est un couple $(C, \pi)$ associant un ensemble de 
critères à une partition des identifiants : 
$C \in \mathscr{P}(\mathcal{C})$ et $\pi \in \Pi(Tid(r))$. Les éléments de ce couple doivent être
liés par les conditions suivantes : $C = f(\pi)$ , $\pi = g(C) = \pi_C$. 
\\
Soit $c_a = (C_{c_a}, \pi_{c_a})$ un concept \accord de $r$, nous appelons $\pi_{c_a}$ 
\emph{l'extension} de $c_a$ (notée $ext(c_a)$) et $C_{c_a}$ son \emph{intension} (notée $int(c_a)$). 
L'ensemble de tous les concepts \accords d'une relation $r$ est noté $\ConceptsAccords(r)$.
\end{definition}

\begin{theorem}[Treillis des concepts \accords]
Soit $\ConceptsAccords(r)$ l'ensemble des concepts \accords d'une relation $r$.
L'ensemble ordonné $\langle \ConceptsAccords(r), \leq\footnotemark\rangle$ forme un 
treillis complet nommé treillis des concepts \accords.
\footnotetext{Soit $(C_1, \pi_1)$, $(C_2, \pi_2) \in \ConceptsAccords(r)$,
$(C_1, \pi_1) \leq (C_2, \pi_2) \Leftrightarrow C_1 \subseteq C_2$ 
(ou de manière équivalente $\pi_2 \sqsubseteq \pi_1$).}
$\forall P \subseteq \ConceptsAccords(r)$, l'\emph{infimum} ou borne inférieure ($\bigwedge$) et 
\emph{supremum} ou borne supérieure ($\bigvee$) sont donnés ci-après :
\begin{align*}
\bigwedge P &= (\bigcap_{c_a \in P} int(c_a),\ h'(\Plus_{c_a \in P} ext(c_a)))        \\
\bigvee P &= (h(\bigcup_{c_a \in P} int(c_a)),\ \Fois_{c_a \in P} ext(c_a))
\end{align*}
\end{theorem}
\begin{proof}
Puisque le couple $gc = (f,\ g)$ est une connexion de Galois, 
le treillis des concepts \accords est un treillis des concepts d'après le théorème 
fondamental de Wille~\citep{book/GanterW99}.
\end{proof}
\begin{example}\label{ex:treillis_concepts_accords}
La figure~\ref{treillis_des_concepts_accords} donne le diagramme de Hasse du treillis des concepts 
\accords de la relation \textsc{Logement}. Le couple $(ECV, \{1,2,35,4\})$ est un concept 
\accord car d'après les exemples précédents on a $g(ECV) = \{1,2,35,4\}$ et 
$f(\{1,2,35,4\}) = ECV$. À l'inverse le couple $(EC, \{1,2,35,4\})$ ne constitue pas un concept 
\accord car $f(\{1,2,35,4\}) \neq EC$. Soit $c_a = (ECV, \{1,2,35,4\})$ et 
$c_b = (PV, \{1,25,3,4\})$ deux concepts \accords, nous avons donc : 
\begin{align*}
c_a \wedge c_b = & (ECV \cap PV ,\ h'( \{1,2,35,4\} \Plus \{1,25,3,4\}) )\\
               = & (V ,\ h'(\{1,235,4\}) ) = (V ,\ \{1,235,4\})\\
c_a \vee c_b   = & (h(ECV \cup PV) ,\ \{1,2,35,4\} \Fois \{1,25,3,4\} )\\
               = & (h(PECV) ,\ \{1,2,3,4,5\}) = (PECV ,\ \{1,2,3,4,5\})\\
\end{align*}
\end{example}
\setlength{\niveauZero}{0bp}
\setlength{\niveauUn}{25bp}
\setlength{\niveauDeux}{50bp}
\setlength{\niveauTrois}{75bp}
\begin{figure}
  \centering
    \begin{tikzpicture}[>=latex,line join=bevel,]
      \node (bottom) at (120bp,\niveauZero) [draw,draw=none] {$(\emptyset, \{12345\})$};
      
      \node (p1)     at (0bp ,\niveauUn) [draw,draw=none] {$(P, \{13,25,4\})$};
      \node (p4)     at (80bp,\niveauUn) [draw,draw=none] {$(V, \{1,235,4\})$};
      \node (p2)     at (160bp,\niveauUn) [draw,draw=none] {$(E, \{12,345\})$};
      \node (p3)     at (240bp,\niveauUn) [draw,draw=none] {$(C, \{1,24,35\})$};

      \node (p6)     at (40bp,\niveauDeux) [draw,draw=none] {$(PV, \{1,25,3,4\})$};
      \node (p5)     at (200bp,\niveauDeux) [draw,draw=none] {$(ECV, \{1,2,35,4\})$};
      
      \node (top)    at (120bp,\niveauTrois) [draw,draw=none] {$(PECV, \{1,2,3,4,5\})$};
      
      \draw [] (p6) -- (top);
      \draw [] (p4) -- (p6);
      \draw [] (bottom) -- (p2);
      \draw [] (bottom) -- (p1);
      \draw [] (bottom) -- (p4);
      \draw [] (bottom) -- (p3);
      \draw [] (p2) -- (p5);
      \draw [] (p5) -- (top);
      \draw [] (p3) -- (p5);
      \draw [] (p4) -- (p5);
      \draw [] (p1) -- (p6);
    \end{tikzpicture}
  \caption{Diagramme de Hasse du treillis des concepts \accords de la relation \textsc{Logement}%
           \label{treillis_des_concepts_accords}}
\end{figure}
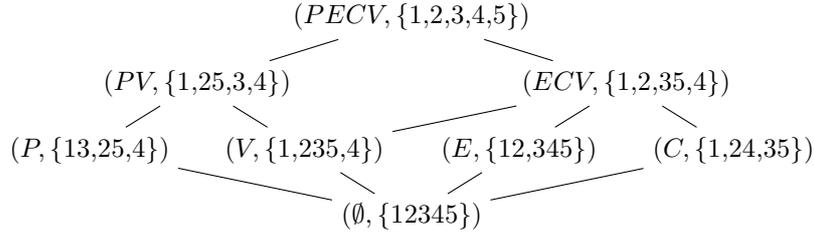
\begin{proposition}\label{prop:partitions_egales}
Pour tout ensemble de critères $C \subseteq \mathcal{C}$, la partition associée $\pi_C$ est 
identique à la partition de sa fermeture. 
\[
\forall C \subseteq \mathcal{C}, \pi_C = \pi_{h(C)}
\]
\end{proposition}
\begin{proof}
Par définition $\forall C \subseteq \mathcal{C}$, $\pi_C = g(C)$ et $h(C) = f(g( C ))$. 
Nous avons donc $\pi_{h(C)}  = g(f(g(C))$. Or, le couple $gc = (f,\ g)$ étant une connexion de Galois, 
on a $g \circ f \circ g  = g$~\citep{book/GanterW99}. Ainsi, on a donc bien 
$\pi_{h(C)}  = g(f(g(C))) = g(C) = \pi_C$.
\end{proof}
\begin{example}
Avec la relation \textsc{Logement}, en prenant comme ensemble de critères $EC$, d'après les 
exemples~\ref{ex:f_g} et~\ref{ex:fermeture}, on a : $\pi_{EC} = g(EC) = \{1,2,35,4\}$ 
et $\pi_{h(EC)} =g(h(EC)) = g(ECV) = \{1,2,35,4\}$.
\end{example}

La proposition précédente signifie que la fermeture d'un ensemble de critères $C$ peut être vue comme 
le plus grand sur-ensemble de $C$ ayant la même partition.

\section{Treillis des Concepts \skylines pour la matérialisation partielle du \skycube}
Le treillis des concepts \skylines est un treillis des concepts \accords contraints. 
Nous démontrons une propriété fondamentale de notre treillis nous permettant de ne matérialiser que 
partiellement le \skycube en éliminant certains cuboïdes. Nous montrons ensuite comment de tels cuboïdes 
peuvent être reconstruits facilement.

\subsection{Treillis des concepts \skylines}


Soit $\pi_C$ une partition de $r$ sur $C$. Par définition, les tuples d'une même classe d'équivalence
$[t]_C \in \pi_C$ sont indistinguables sur $C$ (ils partagent tous la même projection sur $C$).
Si $t$ est non dominé sur $C$, il en sera de même pour tous les autres tuples de sa classe (et 
réciproquement). Il suffit donc de tester la dominance d'un seul tuple d'une classe pour savoir si 
l'ensemble des tuples de la classe appartient ou non au Skycuboïde selon $C$. Ainsi, pour optimiser 
le calcul d'un Skycuboïde selon $C$ à partir de sa partition $\pi_C$, on conserve uniquement 
un seul représentant de chaque classe d'équivalence. Cet ensemble est noté $reps(\pi_C)$. On 
diminue alors la taille de l'entrée en éliminant les tuples qui auraient conduit à un grand nombre 
de comparaisons inutiles. Pour prendre en compte les particularités du calcul de l'opérateur 
\skyline sur une partition, nous introduisons l'opérateur suivant.

\begin{definition}
Soit $C$ un ensemble de critères et $\pi$ une partition de $r$.
On définit un nouvel opérateur $\piSKY$ de la manière suivante :
\begin{align*}
\piSKY_C(\pi_C) &= \{ [t_i] \in \pi_C \mid \forall t_j \in r \text{ on a } t_j \nsucc_C t_i \} \\
                &= \{ [t_i] \in \pi_C \mid t_i \in \SKY_C(r)\}
\end{align*}
\end{definition}

\begin{definition}[Concepts \skylines]
Soit $c_a = (C, \pi) \in \textsc{ConceptsAccords}(r)$ un concept \accord d'une relation $r$.
Le concept \skyline associé $c_s$ est défini de la manière suivante :
\[
c_s = (C, \piSKY_{C}(\pi))
\]
Il y a exactement autant de concepts \skylines que de concepts \accords.
L'ensemble des concepts \skylines associés aux concepts \accords de $r$ est noté 
$\textsc{ConceptsSkylines}(r)$.
\end{definition}

Les concepts \skylines sont des concepts \accords où les partitions ont été contraintes.
Ainsi, ce type de concept n'est plus forcément ordonné par la relation $\sqsubseteq$ entre partitions.
La relation d'ordre $\leq$ entre concepts \skylines s'exprime donc de la manière suivante :
Pour tout $(C_1, \piSKY_{C_1}(\pi_1))$, $(C_2, \piSKY_{C_2}(\pi_2)) \in \ConceptsSkylines(r)$, alors
$(C_1, \piSKY_{C_1}(\pi_1)) \leq (C_2, \piSKY_{C_2}(\pi_2)) \Leftrightarrow C_1 \subseteq C_2$.

\begin{definition}[Treillis des concepts \skylines]
L'ensemble ordonné $\langle \ConceptsSkylines(r), \leq\rangle$ forme un treillis complet nommé 
treillis des concepts \skylines. Il est isomorphe au treillis des concepts \accords.
\end{definition}

\begin{figure}
  \centering
    \begin{tikzpicture}[>=latex,line join=bevel,]
      \node (bottom) at (120bp,\niveauZero) [draw,draw=none] {$(\emptyset, \{12345\})$};
      
      \node (p1)     at (0bp , \niveauUn) [draw,draw=none] {$(P, \{25\})$};
      \node (p4)     at (80bp, \niveauUn) [draw,draw=none] {$(V, \{235\})$};
      \node (p2)     at (160bp,\niveauUn) [draw,draw=none] {$(E, \{345\})$};
      \node (p3)     at (240bp,\niveauUn) [draw,draw=none] {$(C, \{24\})$};

      \node (p6)     at (40bp,\niveauDeux) [draw,draw=none] {$(PV, \{25\})$};
      \node (p5)     at (200bp,\niveauDeux) [draw,draw=none] {$(ECV, \{2,35,4\})$};
      
      \node (top)    at (120bp,\niveauTrois) [draw,draw=none] {$(PECV, \{2,4,5\})$};
      
      \draw [] (p6) -- (top);
      \draw [] (p4) -- (p6);
      \draw [] (bottom) -- (p2);
      \draw [] (bottom) -- (p1);
      \draw [] (bottom) -- (p4);
      \draw [] (bottom) -- (p3);
      \draw [] (p2) -- (p5);
      \draw [] (p5) -- (top);
      \draw [] (p3) -- (p5);
      \draw [] (p4) -- (p5);
      \draw [] (p1) -- (p6);
    \end{tikzpicture}
  \caption{Diagramme de Hasse du treillis des concepts \skylines de la relation \textsc{Logement}%
           \label{treillis_des_concepts_skylines}}
\end{figure}
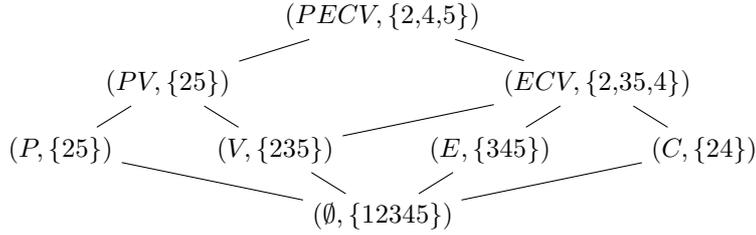

\begin{example}
La figure~\ref{treillis_des_concepts_skylines} donne le diagramme de Hasse du treillis des concepts 
\skylines de la relation \textsc{Logement}. D'après l'exemple~\ref{ex:treillis_concepts_accords},
le couple $c_a = (ECV, \{1,2,35,4\})$ est un concept \accord. Le concept \skyline 
$c_s$ associé est $c_s = (ECV, \piSKY_{ECV}(\{1,2,35,4\})) = (ECV, \{2,35,4\})$. L'identifiant $1$ 
est éliminé de l'extension par l'opérateur \piSKY car le tuple $t_1$ est dominé par $t_2$.
\end{example}

\subsection{Matérialisation partielle du \skycube}

Ce paragraphe propose le treillis des concepts \skylines comme une matérialisation partielle des 
\skycubes. L'idée sous-jacente, afin d'obtenir une représentation réduite, est d'éliminer les 
cuboïdes les plus efficacement reconstructibles.

\begin{definition}[Condition de non accord]
  Soit $r$ une relation et $C$ un ensemble de critères. La condition $\CNA_C(r)$ est vérifiée quand :
  \[
    \nexists t_i,\ t_j \in r \text{ tels que } t_i[C] = t_j[C]
    \text{ avec } i \neq j,\ C \subseteq \mathcal{C}
  \]
  Lorsque $\CNA_C(r)$ est vérifiée, $\CNA_X(r)$ l'est aussi avec $X$ un sur-ensemble de $C$.
\end{definition}

Cette condition de non accord est une version «~affaiblie~» de la condition de valeurs 
distinctes~\citep{vldb/YuanLLWYZ05} puisqu'elle s'applique sur les projections et non pas sur 
les valeurs individuelles de chacun des critères.

\begin{definition}[Dominance sous \CNA]\label{def:cna}
  Soit $C \subseteq \mathcal{C}$ un ensemble de critères et $r$ une relation. La relation de 
  dominance $t \succ_C t$ sous l'hypothèse $\CNA_C(r)$ s'exprime plus simplement :
  \[
    \text{Soit } C \text{ tel que } \CNA_C(r), \ 
    \forall t_i,\ t_j \in r
    \text{ on a } t_j \succ_C t_i
    \text{ ssi } \forall c \in C,\ t_j[c] \leq t_i[c]
    \text{ avec } i \neq j
  \]
\end{definition}

\begin{lemma}\label{lem:skyline_dim_sup}
  Soit $r$ une relation. Pour tout ensemble de critères $C \subseteq \mathcal{C}$ vérifiant l'hypothèse de non accord $\CNA_C(r)$,
  on a $\SKY_C(r) \subseteq \SKY_{C \cup c_0}(r)$ avec $c_0 \in \mathcal{C} \smallsetminus C$.
  \begin{proof}
      Sous l'hypothèse $\CNA_C(r)$, on a :
        $t_i \in \SKY_C (r)$
        $\Rightarrow$  $\forall t_j \in r$  avec $i \neq j$, on a $t_j \nsucc_C t_i$ 
        $\Rightarrow$  $\forall t_j \in r$ avec $i \neq j$, $\exists c \in C$ tel que $t_j[c] > t_i[c]$ (\text{\emph{cf.} définition~\ref{def:cna}})
        $\Rightarrow$  $\forall t_j \in r$ avec $i \neq j$, $\exists c \in C \cup c_0$ tel que  $t_j[c] > t_i[c]$
        $\Rightarrow$  $\forall t_j \in r$ avec $i \neq j$, on a $t_j \nsucc_{C \cup c_0} t_i$
        $\Rightarrow$  $t_i \in \SKY_{C \cup c_0} (r)$.
      Sous l'hypothèse $\CNA_C(r)$, on a donc $\SKY_C (r) \subseteq \SKY_{C \cup c_0} (r)$.
  \end{proof}
\end{lemma}

Le contre exemple suivant montre que la propriété réciproque n'est pas vérifiée.

\begin{example}\label{ex:non_egalite_skylines}
  Soit $r = \{ t_1, t_2 \}$ une relation (avec $t_1=(0,\ 1)$ et $t_2=(1,\ 0)$) et 
  $\mathcal{C} = \{ A, B \}$ l'ensemble de ses critères. La condition $CNA_A(r)$ est bien vérifiée.
  On a $t_2 \notin \SKY_A(r)$ car $t_1 \succ_A t_2$ alors que $t_2$ appartient bien à $\SKY_{A \cup B}(r)$.
\end{example}

\begin{cor}
Soit $C$ tel que $\CNA_C(r)$ est vérifiée.
Par définition, pour tout $X$ ($C\subseteq X$) vérifiant la condition $\CNA_X(r)$, on a $\SKY_C(r) \subseteq \SKY_X(r)$.
\end{cor}

Cette propriété est satisfaite pour tout sur-ensemble de $C$ jusqu'à l'ensemble de tous les critères $\mathcal{C}$.

\begin{theorem}[Théorème fondamental]\label{theoreme_fondamental}
Soit $r$ une relation, $C$ un ensemble de critères et $h(C)$ sa fermeture. Alors :
\[
\forall C \subseteq \mathcal{C} \text{ on a } \SKY_C(r) \subseteq \SKY_{h(C)}(r)
\]
\end{theorem}

\begin{proof}
Par définition, $t_i \in \SKY_C(r) \text{ ssi } \exists [t] \in \piSKY_C(\pi_C) \text{ telle que } i \in [t]$.
On cherche donc à montrer que $\piSKY_C(\pi_C) \subseteq \piSKY_C(\pi_{h(C)})$.
On sait que pour tout $X$ tel que $C \subseteq X \subseteq h(C)$ on a $\pi_X = \pi_{C}$.
Soit $E = \{ t_i \in r \mid i \in reps(\pi_{C}) \}$ l'ensemble des tuples représentants de cette 
partition. On ne peut se baser sur $E$ pour le calcul des \skylines sur $X$ que dans l'intervalle 
$C \subseteq X \subseteq h(C)$, autrement les classes d'équivalences n'étant plus égales il est 
impossible de garantir une reconstruction correcte du \skyline total à partir des représentants 
des classes d'équivalences. Aussi, la condition $\CNA_X(E)$ est vérifiée car chaque tuple 
représentant est distinguable d'un autre donc d'après le lemme~\ref{lem:skyline_dim_sup} on a
$\SKY_C(E) \subseteq \SKY_{h(C)}(E)$ d'où $\SKY_C(r) \subseteq \SKY_{h(C)}(r)$.
\end{proof}

\begin{example}
Dans la relation \textsc{Logement}, on a $\SKY_{EC}($\textsc{Logement}$) = \{ t_4 \}$.
De plus, $h(EC) = ECV$ et $\SKY_{ECV}($\textsc{Logement}$) = \{ t_2, t_3, t_5, t_4 \}$.
$\SKY_{EC}($\textsc{Logement}$) \subseteq \SKY_{h(EC)}($\textsc{Logement}$)$ est donc bien vérifiée.
Aussi, on a $\SKY_E($\textsc{Logement}$) = \{ t_3, t_4, t_5 \}$ avec $h(E) = E$.
On note donc la non inclusion $\SKY_{E}($\textsc{Logement}$) \nsubseteq \SKY_{EC}($\textsc{Logement}$)$.
\end{example}

Le théorème précédent montre une inclusion entre certains cuboïdes. Plus exactement, pour tout 
ensemble de critères, il y a une chaîne d'inclusions allant de ses générateurs minimaux jusqu'à 
sa fermeture. Cela implique que l'on peut calculer un cuboïde à partir d'un autre qui le contient.
Ainsi au lieu d'utiliser la relation complète, on ne considère qu'un sous-ensemble restreint de 
celle-ci. Les concepts \skylines représentent les plus grands cuboïdes (selon l'inclusion) 
permettant d'en calculer d'autres. Ainsi en ne conservant que les concepts \skylines (\emph{i.e.} 
non matérialisation des cuboïdes non fermés), on peut reconstruire rapidement les cuboïdes manquants 
simplement en trouvant le fermé à partir duquel ils peuvent être recalculés.

De plus grâce à la notion de classe d'équivalence des concepts \skylines, on évite un grand 
nombre de comparaisons inutiles. Les éléments indistinguables n'étant considérés que par groupes, 
la complexité du calcul ne dépend plus du nombre de tuples mais du nombre de 
groupes (classes d'équivalence).

Le treillis des concepts \skylines constitue donc une matérialisation partielle du \skycube, où 
l'information non matérialisée est efficacement recalculable.



\section{Comparaison avec les travaux antérieurs}
L'opérateur \skyline~\citep{icde/BorzsonyiKS01} s'intéresse à trouver les éléments les plus 
intéressants dans un contexte base de données. Il tire ses origines du 
\emph{maximal vector problem}~\citep{tpa/barndorffS66,jacm/BentleyKST78}. Cependant les particularités 
du contexte font que des algorithmes spécifiques ont été développés. Les plus connus 
d'entre eux sont \textsc{Bnl}~\citep{icde/BorzsonyiKS01}, \textsc{Sfs}~\citep{icde/ChomickiGGL03}, 
\textsc{Less}~\citep{vldbj/GodfreySG07}. D'autres algorithmes plus efficaces tels que 
\textsc{Bbs}~\citep{tods/PapadiasTFS05} ont été proposés mais se basent sur des structures d'index
coûteuses à maintenir.

Le \skyline étant un outil d'aide à la décision, l'utilisateur va généralement calculer plusieurs 
\skylines avant de trouver celui qui l'intéresse vraiment. Pour répondre à cette problématique, le 
concept de \skycube~\citep{vldb/YuanLLWYZ05,vldb/PeiJET05} a été proposé. L'idée générale étant de 
pré-calculer tous les \skylines, il est impératif de restreindre autant que possible le coût de 
stockage du résultat.

Une première approche de réduction a été proposée par~\cite{tods/PeiYLJELWTYZ06}.
Son objectif est de résoudre le problème de l'appartenance d'un objet donné à différents Skycuboïdes.
Sa solution consiste à proposer une représentation du \skycube fondée sur l'analyse formelle des 
concepts où chaque nœud correspond à un couple composé d'une classe d'équivalence (ensemble d'objets) 
et des sous-espaces dans lesquels elle est \skyline. Le principal défaut de cette approche orientée 
valeur est que le nombre de nœuds est borné par la cardinalité du treillis des parties des tuples 
($|\mathscr{P}(Tid(r))|$). Les objets \skylines sont considérés avant tout, ce qui fait que pour 
reconstruire un Skycuboïde il faut parcourir un grand nombre de nœuds du treillis. 
Dans les approches orientées attribut comme la nôtre, le nombre de nœuds est bien plus restreint 
car il est borné par $|\mathscr{P}(\mathcal{C})|$.

Une autre approche intéressante est celle de~\cite{sigmod/XiaZ06}.
Elle est orientée cuboïde (orientée attribut).
L'objectif ici est de proposer une réduction très importante du \skycube en éliminant les éléments
considérés comme redondants des différents cuboïdes. 
Son inconvénient est que la reconstruction d'un Skycuboïde est délicate et relativement coûteuse car
il faut parcourir une structure de données pour retrouver les éléments redondants.
De plus, contrairement à l'approche de~\cite{tods/PeiYLJELWTYZ06,icde/PeiFLW07} et à la nôtre,
celle-ci n'est pas fondée sur un support théorique aussi solide que l'analyse des concepts formels.
Cependant, l'un des principaux intérêts de cette approche en plus de la réduction importante du coût 
de stockage est l'efficacité de la mise à jour des données.

Notre approche ayant pour objectif de reconstruire les Skycuboïdes à moindre coût, elle se comporte
idéalement dans ce cas-là. Malgré cette orientation ciblée, elle représente un bon compromis que ce soit 
pour la mise à jour ou pour la réduction de l'espace de stockage.
\section{Conclusion}
Dans ce papier, nous avons proposé le treillis des concepts \skylines qui est une instance contrainte
d'un cadre formel plus général : le treillis des concepts \accords. Cette nouvelle structure 
permet non seulement la matérialisation partielle sans perte d'information mais aussi la 
reconstruction efficace des cuboïdes manquants. 
Nous pouvons facilement étendre notre matérialisation au cas des $\epsilon$-\skylines ou 
\skylines approximatifs~\citep{tods/PapadiasTFS05},
en généralisant la définition d'ensemble en accord en remplaçant l'égalité stricte par une égalité à 
$\epsilon$ près. Une telle extension permet à l'utilisateur de relâcher la contrainte de dominance 
lorsque le nombre de résultats n'est pas suffisant.

Nous travaillons actuellement sur les aspects algorithmiques aussi bien pour le calcul du treillis 
des concepts \skylines que pour la reconstruction des résultats. Ce travail a pour objectif de valider 
la contribution théorique décrite dans cet article.
\bibliographystyle{rnti}
\bibliography{biblio_propre}
\appendix
\section{Treillis des partitions}
\begin{definition}[Partition d'un ensemble]
  Soit E un ensemble, une partition $\pi(E)$\footnote{S'il n'y a pas d'ambigüité sur l'ensemble $E$
  nous notons la partition $\pi$} d'un ensemble $E$ est une famille de parties de cet ensemble
  telle que chaque élément de $E$ appartient exactement à une seule de ces familles (ou classes). 
  Autrement dit, $\pi(E)$ est une famille d'ensembles deux à deux disjoints ($\forall X, Y \in \pi(E)$ 
  on a $X \cap Y = \emptyset$) dont l'union est égale à $E$ ($\bigcup_{X \in \pi(E)} = E$).
\end{definition}
\begin{definition}[Relation d'ordre entre partitions]
Soit $\pi(E)$, $\pi'(E)$ deux partitions d'un même ensemble $E$, $\pi(E)$ est dite plus fine que $\pi'(E)$
si et seulement si toute classe de $\pi(E)$ est obtenue par subdivision de classes de $\pi'(E)$
\footnote{De façon équivalente, $\pi(E)$ est plus fine que $\pi'(E)$ si et seulement si toute classe 
de $\pi'(E)$ est l'union de classes de $\pi(E)$.}. La relation de finesse entre deux 
partitions est une relation d'ordre partiel notée $\sqsubseteq$. Elle est définie comme suit :
\[
\pi(E) \sqsubseteq \pi'(E) \Leftrightarrow \pi(E) \text{ est plus fine que } \pi'(E)
\footnote{Réciproquement $\pi'(E)$ est dite plus grossière que $\pi(E)$.}
\Leftrightarrow (\forall X \in \pi(E),\ \exists X' \in \pi'(E),\ X \subseteq X' )
\]
\end{definition}
\begin{definition}[Produit de partitions] 
Soit $\pi(E)$ et $\pi'(E)$ deux partitions d'un même ensemble $E$. 
Le produit des partitions $\pi(E)$ et $\pi'(E)$, noté $\pi(E) \bullet \pi'(E)$, est obtenu comme suit :
\[
\pi(E) \bullet \pi'(E) = 
\{ Z = X \cap Y \mid Z\neq\emptyset,\ X \in \pi(E) \text{ et } Y \in \pi'(E)\}
\]
\end{definition}
Avant de définir l'opérateur somme entre deux partitions, nous définissons une fonction 
outil $R$ : 
\[
   R(e,F) = \bigcup_{\substack{X \in F \\ e \in X}} X
\]
avec $e$ un élément d'un ensemble $E$, $F$ une famille de parties de $E$. $R(e,F)$ correspond à 
l'union des ensemble de $F$ contenant l'élément $e$.
\begin{definition}[Somme de partitions] 
Soit $\pi(E)$ et $\pi'(E)$ deux partitions d'un même ensemble $E$. La somme des partitions $\pi(E)$
et $\pi'(E)$, notée $\pi(E) + \pi'(E)$, est obtenue par fermeture transitive de l'opération qui fait 
correspondre à un élément de $E$ l'ensemble des éléments de ses classes dans $\pi(E)$ et 
$\pi'(E)$~\citep{book/Birkhoff70}. La suite $S$ est définie ci-dessous pour formaliser ce calcul : 
\[
  \left\{
  \begin{array}{lcl}
    S_0 &=&\displaystyle\max_{\subseteq}({\pi(E) \cup \pi'(E)})\\
    S_n &=&\displaystyle\max_{\subseteq}(\{ R(e, S_{n-1}) \mid e \in E \})
  \end{array}\right.
\]
Ainsi l'opérateur somme peut se définir de la manière suivante :
\[
  \pi(E) + \pi'(E) = S_k \text{ avec $k$ tel que } S_k = S_{k-1}
\]
\end{definition}
\begin{theorem}[Treillis des partitions]
Soit $\Pi(E)$ l'ensemble des partitions possibles d'un ensemble $E$. L'ensemble ordonné 
$\langle\Pi(E), \sqsubseteq \rangle$ forme un treillis complet nommé treillis des partitions de $E$.
$\forall P \subseteq \Pi(E)$, son \emph{infimum} ou borne inférieure ($\bigwedge$) et son \emph{supremum} 
ou borne supérieure ($\bigvee$) sont donnés ci-après :
\[
  \bigwedge P = \Fois_{\pi \in P}~\pi,  \hspace{3cm}  \bigvee P = \Plus_{\pi \in P}~\pi 
\]
\end{theorem}
\setlength{\niveauZero}{0bp}
\setlength{\niveauUn}{25bp}
\setlength{\niveauDeux}{75bp}
\setlength{\niveauTrois}{100bp}
\begin{figure}
  \centering
  \begin{tikzpicture}[>=latex,line join=bevel,]
  
    \node (bottom) at (150bp,\niveauZero) [draw,draw=none] {\{1234\}};

    \node (p6)  at (0bp  ,\niveauUn) [draw,draw=none] {\{13,24\}};  
    \node (p1)  at (50bp ,\niveauUn) [draw,draw=none] {\{1,234\}};
    \node (p2)  at (100bp,\niveauUn) [draw,draw=none] {\{2,134\}};
    \node (p3)  at (150bp,\niveauUn) [draw,draw=none] {\{3,124\}};
    \node (p5)  at (200bp,\niveauUn) [draw,draw=none] {\{12,34\}};    
    \node (p4)  at (250bp,\niveauUn) [draw,draw=none] {\{4,123\}};
    \node (p7)  at (300bp,\niveauUn) [draw,draw=none] {\{14,23\}};

    \node (p9)  at (0bp  ,\niveauDeux) [draw,draw=none] {\{1,3,24\}};
    \node (p8)  at (60bp ,\niveauDeux) [draw,draw=none] {\{1,2,34\}};
    \node (p12) at (120bp,\niveauDeux) [draw,draw=none] {\{2,4,13\}};
    \node (p11) at (180bp,\niveauDeux) [draw,draw=none] {\{2,3,14\}};
    \node (p13) at (240bp,\niveauDeux) [draw,draw=none] {\{3,4,12\}};        
    \node (p10) at (300bp,\niveauDeux) [draw,draw=none] {\{1,4,23\}};

    \node (top) at (150bp,\niveauTrois) [draw,draw=none] {\{1,2,3,4\}};

    \draw [-] (p4) -- (p13);
    \draw [-] (bottom) -- (p1);
    \draw [-] (p7) -- (p10);
    \draw [-] (bottom) -- (p5);
    \draw [-] (p13) -- (top);
    \draw [-] (p4) -- (p10);
    \draw [-] (p6) -- (p9);
    \draw [-] (p8) -- (top);
    \draw [-] (p6) -- (p12);
    \draw [-] (bottom) -- (p4);
    \draw [-] (p3) -- (p11);
    \draw [-] (p10) -- (top);
    \draw [-] (bottom) -- (p3);
    \draw [-] (p1) -- (p9);
    \draw [-] (p2) -- (p11);
    \draw [-] (p5) -- (p13);
    \draw [-] (bottom) -- (p7);
    \draw [-] (p2) -- (p8);
    \draw [-] (p1) -- (p10);
    \draw [-] (p3) -- (p9);
    \draw [-] (p9) -- (top);
    \draw [-] (p4) -- (p12);
    \draw [-] (bottom) -- (p2);
    \draw [-] (p11) -- (top);
    \draw [-] (p7) -- (p11);
    \draw [-] (p1) -- (p8);
    \draw [-] (p2) -- (p12);
    \draw [-] (bottom) -- (p6);
    \draw [-] (p3) -- (p13);
    \draw [-] (p5) -- (p8);
    \draw [-] (p12) -- (top);
    
    \node[right=1em] at (p7.east)  (l3) { };
    \node at (l3 |- bottom) (l4){ };
    \node at (l3 |- top) (l1){ };
    \draw[->,black,line width=1pt] (l1) to (l4) node [midway,right] {$\sqsubseteq$};
  \end{tikzpicture}
  \caption{Diagramme de Hasse du treillis des partitions de l'ensemble $E = \{1, 2, 3, 4\}$%
           \label{treillis_des_partitions}}
\end{figure}
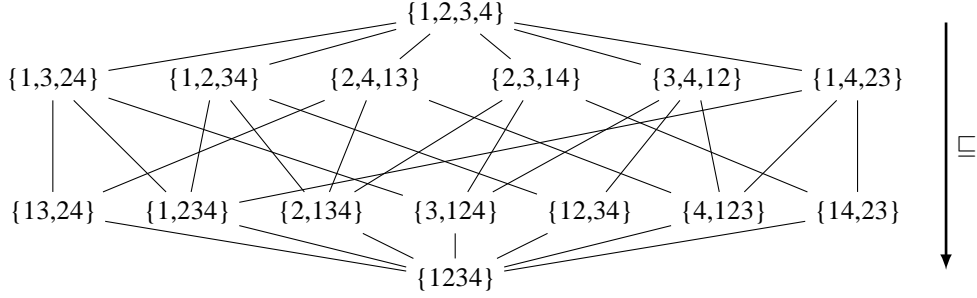
\begin{example}
Le diagramme de Hasse du treillis des partitions de l'ensemble $E = \{1, 2, 3, 4\}$ est donné 
dans la figure~\ref{treillis_des_partitions}. Par soucis d'uniformité avec le treillis des concepts
\accords, le treillis est représenté inversé par rapport à sa disposition naturelle. 
Les partitions les plus grossières sont en bas et les plus fines en haut.
\end{example}
\end{document}